\documentclass[reprint,amsmath,amssymb,aps,prx,tightenlines,nobibnotes,superscriptaddress,longbibliography]{revtex4-2}

\usepackage{verbatim}
\usepackage{tikz}
\usetikzlibrary{matrix,fit}

\usepackage{graphicx}
\usepackage{float}
\usepackage{dcolumn}
\usepackage{bm}
\usepackage{xcolor}
\usepackage{color}
\usepackage{cancel}
\usepackage{natbib}
\usepackage{setspace}
\usepackage{pifont}
\usepackage{mathtools}
\usepackage{layouts}
\usepackage{empheq}
\usepackage{multirow}
\usepackage{amsmath}              % Mathematikbefehle
\usepackage{amssymb}              % Unter anderem schöne Mengensymbole
\usepackage{amsfonts}             % Weitere Symbole für equation Umgebung
\usepackage{siunitx}     % Einheiten per Befehl einfügen
\usepackage[applemac]{inputenc}
\usepackage[english]{babel}
\newcommand{\grey}[1]{\textcolor{gray}{#1}}
\newcommand{\pec}{\text{P\'{e}clet}}
\newcommand{\pe}{\ensuremath{\mathrm{Pe}}}
\newcommand{\grayzero}{{\color{lightgray}0}}
\newcommand{\eps}{\ensuremath{\varepsilon}}

\newcommand{\mmat}{\ensuremath{\mathcal{M}}}
\newcommand{\amat}{\ensuremath{\mathcal{J}}}
\newcommand{\bigo}{\ensuremath{\mathcal{O}}}

\begin{document}

\title{Active Fluid Patterning in Inhomogeneous Environments}

\author{Douglas MacMyn Brown}
\affiliation{Rudolf Peierls Centre for Theoretical Physics, Department of Physics, University of Oxford, Parks Road, Oxford OX1 3PU, United Kingdom}
\author{Alexander Mietke}
\email{alexander.mietke@physics.ox.ac.uk}
\affiliation{Rudolf Peierls Centre for Theoretical Physics, Department of Physics, University of Oxford, Parks Road, Oxford OX1 3PU, United Kingdom}

\begin{abstract}
Active stresses in biological cells and tissues drive many developmental processes. However, increasing experimental evidence suggests that additional mechanical interactions with surrounding material can play a crucial role in guiding these processes. We introduce a minimal model of this scenario and investigate how pattern formation in an active material can be controlled by an inhomogeneous environment. Specifically, we consider an active fluid in which a chemical species regulates local active stresses and is redistributed by the resulting flows. We show that active stress patterns within such a fluid exhibit frictiotaxis and systematically characterize how inhomogeneous external friction affects mechanochemical pattern formation instabilities. We find that hydrodynamic screening plays a crucial role in mediating the cross-talk between friction patterns and active fluid self-organization and identify a mechanochemical frustration mechanism that gives rise to pattern oscillations caused by inhomogeneous friction. 
\end{abstract}

\maketitle

\setlength{\parindent}{0pt}

\section{Introduction}\label{sec:introduction}

The interplay of biochemical stress regulation and mechanical properties is crucial for the developmental dynamics of cell and tissues~\cite{lecuitCellSurfaceMechanics2007,mammotoMechanicalControlTissue2010,salbreuxActinCortexMechanics2012,wennekampSelforganizationFrameworkSymmetry2013,hannezoMechanochemicalFeedbackLoops2019,baillesMechanochemicalPrinciplesSpatial2022}. Many morphogenetic processes have been identified in recent years where this interplay is complemented by crucial mechanical interactions of living matter with its complex surroundings. For example, interactions between cells and the extracellular matrix (ECM) or enclosing egg shells facilitate the elongation, flow and fusion of biological tissues~\hbox{\cite{kellerMechanismsElongationEmbryogenesis2006,munsterAttachmentBlastodermVitelline2019,goodwinBasalCellExtracellularMatrix2016}}. 
% The corresponding mechanical and biochemical interactions can be mediated directly, or through more complex signaling pathways via intermediate tissues~\cite{gillardForceTransmissionThree2019}. 
Interactions with surroundings that provide a rigid substrate can affect spindle orientation during cell division~\cite{theryExtracellularMatrixGuides2005}, enable division of \textit{Dictyostelium}~\cite{tairaNovelModeCytokinesis2017}, and facilitate cell migration~\cite{leberreGeometricFrictionDirects2013, shellardFrictiotaxisUnderliesFocal2025, sarkarCellMigrationDriven2020,pagesCellClustersAdopt2022, liuConfinementLowAdhesion2015,moreauIntegratingPhysicalMolecular2018, karlingImmuneCellsAdapt2025} which is linked to cancer metastasis~\cite{thullbergOncogenicHRasV122007, byunCharacterizingDeformabilitySurface2013}. While mechanochemical organization principles of these processes have been studied in both theoretical \cite{boisPatternFormationActive2011, kumarPulsatoryPatternsActive2014,mooreSelfOrganizingActomyosinPatterns2014, veermanTuringFarfromequilibriumPatterns2021,brinkmannPostTuringTissuePattern2018, mietkeSelforganizedShapeDynamics2019} and experimental settings \cite{grossGuidingSelforganizedPattern2019, hannezoCorticalInstabilityDrives2015,caldarelliSelforganizedTissueMechanics2024, goehringPolarizationPARProteins2011, arslanAdhesioninducedCorticalFlows2024}, little is known about how external forces contribute to the guidance of emerging patterns in these systems. 

Cells and tissues are non-equilibrium materials that can spontaneously flow and deform due their ability to generate active stresses~\cite{marchettiHydrodynamicsSoftActive2013,julicherHydrodynamicTheoryActive2018,joannyActiveGelsDescription2009}. Their dynamics is often dominated by large-scale deformations at timescales much slower than that of the components~\cite{ranftFluidizationTissuesCell2010}. As a result, hydrodynamic models have been successful in modeling cells and tissues as active fluids~\cite{etournayInterplayCellDynamics2015, copenhagenTopologicalDefectsPromote2021, pallaresStiffnessdependentActiveWetting2022, munsterAttachmentBlastodermVitelline2019, romeoLearningDevelopmentalMode2021, dyeSelforganizedPatterningCell2021, pimpaleCellLineagedependentChiral2020}. Mechanochemical feedback can be included in these models via a stress-regulating chemical species which redistributes itself by the advective flows it generates~\cite{mayerAnisotropiesCorticalTension2010,boisPatternFormationActive2011, kumarPulsatoryPatternsActive2014,mietkeSelforganizedShapeDynamics2019, bonatiRoleMechanosensitiveBinding2022}. The resulting spontaneous symmetry breaking and mechanochemical pattern formation have typically been studied in simple, homogeneous environments.

\begin{figure*}
    \includegraphics{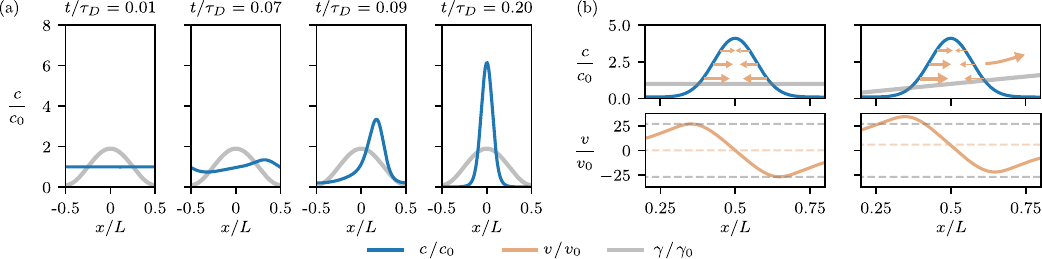}
    \caption{Mechanism of friction-guided contractility localization. (a)~A region of high stress regulator concentration forms spontaneously and aligns with the friction maximum. ($\pe{} = 800, \ell_h/L = 1/ 2\pi, \varepsilon=0.9$) (b)~Gradients in friction $\gamma(x)$ lead to asymmetric flows into contractile patches (right) compared to homogeneous friction (left). As a result, contractile patches move up friction gradients and eventually localize at friction maxima. Gray dashed lines indicate maximum/minimum flow velocities for the homogeneous friction case. Orange dashed lines show average flow velocity across the domain. 
    } \label{fig:time_series_and_v_init_diagram}
\end{figure*}

Here, we study the patterning dynamics of a self-organized active fluid that experiences an inhomogeneous external friction. We find that inhomogeneous friction can pin active stress patterns and show that critical instability thresholds are determined by a competition between the hydrodynamic screening length and the length scale of friction inhomogeneities. We use these insights to explain oscillatory patterns that originate from a mechanism of mechanochemical frustration.

\section{Model}\label{sec:model}
We consider a minimal model of a self-organized active fluid on a periodic domain $x\in[0,L]$ with stress $\sigma$ described by the constitutive relation~\cite{boisPatternFormationActive2011,hannezoCorticalInstabilityDrives2015}
\begin{equation} \label{eq:stress}
    \sigma = \eta \partial_x v + \xi f(c).
\end{equation}
Here, $\eta$ denotes the fluid viscosity and $v$ the flow velocity. The coefficient $\xi$ sets the magnitude of an active contractility which is controlled by a stress-regulating chemical species with concentration $c$.  Following previous work~\cite{ mietkeSelforganizedShapeDynamics2019, mayerAnisotropiesCorticalTension2010}, we consider $f(c)=\frac{c}{1+c}$ which captures the increase and saturation of active stress with increasing regulator concentration.

The stress regulator dynamics are described by the continuity equation
\begin{equation}\label{eq:conservation}
    \partial_t c = D \partial^2_x c - \partial_x(vc) - r(c-c_0),
\end{equation}
with diffusion coefficient $D$ and degradation rate $r$. A homogeneous reference concentration $c_0$ is defined by the recruitment rate $r_{\text{on}}=rc_0$. 
We consider a scenario in which the internal stress is balance by external friction $f^{\text{ext}}=-\gamma v$, giving a force balance
\begin{equation} \label{eq:force-balance}
    \partial_x \sigma = \gamma v.
\end{equation}
To emulate spatially inhomogeneous interactions with the environment, we include a friction $\gamma(x) = \gamma_0 g(x)$ where $\gamma_0$ is an average friction and $g(x)$ is given by
\begin{equation} \label{eq:fric-function}
    g(x) = 1 + \eps \cos\left(\frac{2 \pi  n x}{L}\right).
\end{equation}
This leads to friction variations with a length scale $\ell_f\sim L/n$ ($n\in \{1,2,\ldots\}$) and magnitude~$\eps$.

Throughout this work, we use the system size $L$ as a characteristic length, the diffusion time scale \hbox{$\tau_D=L^2/D$} as a characteristic time and $c_0$ as a characteristic concentration. A characteristic velocity is accordingly defined by $v_0 = D/L$. 
We also define the \pec{} number, describing the ratio of the diffusive timescale $\tau_D$ to the activity-driven advection timescale~$\tau_{\text{adv}}=\eta/\xi$
\begin{equation} \label{eq:peclet-def}
    \pe = \frac{\tau_D}{\tau_\text{adv}} =\frac{\xi L^2}{\eta D}.
\end{equation}

A crucial remaining independent parameter defined by Eq.~(\ref{eq:force-balance}) is the relative hydrodynamic screening length $\ell_h/L$, where $\ell_h=\sqrt{\eta/\gamma_0}$. In an active fluid, $\ell_h$ corresponds to a length scale over which flows generated by a local active force density exponentially decay~\cite{mayerAnisotropiesCorticalTension2010}.

Previous work has considered \mbox{Eqs~(1--5)} with homogeneous friction and without degradation ($r=0$), and has identified a critical \pec{} number, $\pe{}^{*}$, beyond which spatial patterns of stress regulator concentration form spontaneously~\cite{boisPatternFormationActive2011, mooreSelfOrganizingActomyosinPatterns2014}. At steady state, these patterns are stabilized by a balance of active contractile fluxes and diffusive fluxes in and out of local maxima of the stress regulator. We describe in the following how inhomogeneous external friction alters this balance, leads to frictiotaxis and to frustrated oscillatory states when pattern and friction length scales become incommensurable. 

\begin{figure*}[tb]
    \centering
    \includegraphics{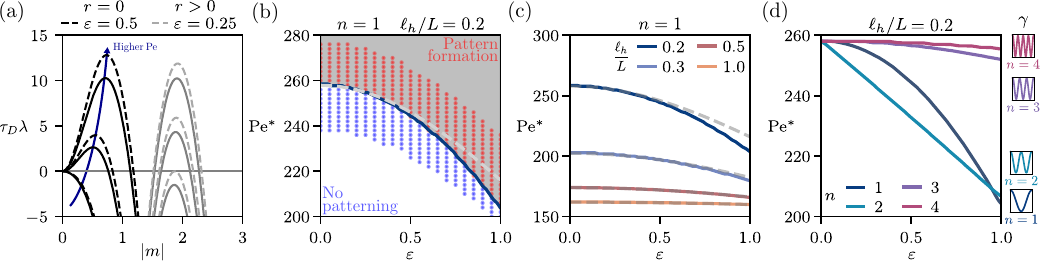}
\caption{Modulations of patterning instabilities due to inhomogeneous friction. (a) Dispersion relation with ($r>0$, gray lines) and without ($r=0$, black lines~\cite{boisPatternFormationActive2011}) degradation. Solid lines show the case of homogeneous friction [$\eps=0$, Eq.~\eqref{eq:homog-dispersion-relation}]. Dashed lines show approximate dispersion obtained using perturbation theory with inhomogeneous friction ($\eps\ne0$, friction pattern Eq.~(\ref{eq:fric-function}) with $n=1$, see~\cite{SI}). Black curves: $r=0, \ell_h/L=1/2 \pi$; gray curves: $r=120,\ell_h/L=1/4\pi$. (b)~Critical \pec{} number $\pe{}^*$ as a function of friction inhomogeneity magnitude~$\eps$. Solid line shows numerically exact prediction, dashed line shows approximation obtained from perturbation theory, Eq.~\eqref{eq:pe-threshold-n1}. Numerical simulation results of the full model with random initial conditions are shown as blue and red dots. (c)~Critical \pec{} number for different hydrodynamic length scales $\ell_h/L$. Dashed lines show predictions from perturbation analysis Eq.~\eqref{eq:pe-threshold-n1}. \pe{}$^*$ becomes independent of $\eps$ when the length scale of hydrodynamic screening $\ell_h$ becomes larger than the friction pattern length scale $\ell_f\sim L$. (d)~Critical \pec{} number for different friction pattern length scales $\ell_f\sim L/n$ show non-monotonic dependence of \pe{}$^*$ on $n$. Figures (b)--(d) use $r=0$ (no degradation).}
    \label{fig:dispersion_relation_and_stability}
\end{figure*}

\section{Results}\label{sec:results}
\subsection{General mechanism of stress regulator frictiotaxis} \label{sec:A:frictiotaxis}
To build intuition for the effect of inhomogeneous friction on this system, we first consider a scenario without stress regulator turnover (\mbox{$r=0$}) in Eq.~(\ref{eq:conservation}).

In a parameter regime where mechanochemical patterns form spontaneously, numerical solutions of Eqs~(\ref{eq:conservation}) and (\ref{eq:force-balance}) suggest that maxima of the stress regulator concentration align with friction maxima, as illustrated in Fig.~\ref{fig:time_series_and_v_init_diagram}a. The underlying mechanism for this type of frictiotaxis is illustrated in Fig.~\ref{fig:time_series_and_v_init_diagram}b. For homogeneous friction ($\eps=0$), Eqs.~\eqref{eq:conservation} and~\eqref{eq:force-balance} are invariant under $(x,v)\rightarrow (-x,-v)$. Under this transformation, concentration gradients $\partial_xc$ switch sign, and any symmetry in concentration gradients is also present in the advective flows. For the contractile fluid we consider here ($\xi>0$), a symmetric region of high stress regulator concentration therefore experiences equal inward flows from both sides~(Fig.~\ref{fig:time_series_and_v_init_diagram}b, left). A small friction gradient breaks that symmetry and leads to reduced (increased) flow into regions of higher (lower) friction (Fig.~\ref{fig:time_series_and_v_init_diagram}b, right). Contractile regions of high regulator concentration therefore translate up the friction gradient, which eventually localizes contractile patches at friction maxima. We note that this phenomenon is distinct from adhesion-independent cellular migration and frictiotaxis~\cite{shellardFrictiotaxisUnderliesFocal2025, bergertForceTransmissionAdhesionindependent2015} where, instead, whole cells move via retrograde contractile flows confined by moving boundaries.

\subsection{Wavelength control of active fluid instabilities}\label{sec:results:degradmodtun}
With inhomogeneous friction, we expect an interplay between the friction pattern length scale \hbox{$\ell_f\sim L/n$} (see Eq.~\ref{eq:fric-function}) and the intrinsic length scales selected by mechanochemical instabilities. The latter can be controlled by the degradation rate $r$ in Eq.~(\ref{eq:conservation}). Formally, degradation makes all homogeneous perturbations unstable and turns the long wavelength type-II instability of the conservative dynamics into potentially short wavelength type-I instabilities~\cite{crossPatternFormationDynamics2009} (see Fig.~\ref{fig:dispersion_relation_and_stability}a). Intuitively, the presence of diffusion and degradation leads to a pattern decay length \smash{$\ell_D\sim\sqrt{D/r}$} that decreases with increasing degradation rate and -- in tandem with hydrodynamically screened active flows~\cite{mietkeMinimalModelCellular2019} -- favors shorter-wavelength patterns. To see this, we linearize Eq.~\eqref{eq:conservation} with flows determined by Eq.~(\ref{eq:force-balance}) for $\eps=0$ around the homogeneous state using an ansatz~\mbox{$c = c_0 + \delta c$}, with \mbox{$\delta c = \sum_{m\in\mathbb{Z}}\delta c_me^{i k_mx + \lambda t}$} \mbox{($k_m = 2 \pi m/L$)}, giving the dispersion relation
\begin{equation} \label{eq:homog-dispersion-relation}
    \lambda(k_m) = Dk_m^2 \left( \frac{c_0 f'(c_0) }{(1 + k_m^2\ell_h^2)} \left ( \frac{\ell_h}{L}\right )^2 \pe -1\right) - r.
\end{equation}
For $r=0$, this recovers the result from Ref.~\cite{boisPatternFormationActive2011} (Fig.~\ref{fig:dispersion_relation_and_stability}a, solid black lines). Modes with the long (system-size) wavelength always become unstable first and are later complemented by secondary, smaller-wavelength instabilities as the P\'eclet number increases. However, as anticipated by the arguments above, Eq.~(\ref{eq:homog-dispersion-relation}) shows that -- for positive degradation rate ($r>0$) and sufficiently small hydrodynamic screening length -- the initial mechanochemical patterning instability can occur at shorter wavelengths (Fig.~\ref{fig:dispersion_relation_and_stability}a, solid gray line).

\subsection{Pattern formation with inhomogeneous friction}\label{results:basic-results-n1}

We first consider the case with no degradation ($r=0$), where intrinsic mechanochemical instabilities are dominated by modes with system-size length scale, and study the impact of a friction pattern $g(x)$ given by Eq.~(\ref{eq:fric-function}) with $n=1$. Despite maintaining the same average friction, we find that friction inhomogeneities lead to mode coupling that can significantly alter the critical P\'eclet number~$\pe^*$. A degenerate perturbation analysis reveals generally increased growth rates (illustrated for $\eps=0.5$ in Fig.~\ref{fig:dispersion_relation_and_stability}a, dashed lines). The modified Jacobian of the linearized system can be expanded to lowest order in \eps{}, which in turn leads to a modified critical \pec{} number given by~\cite{SI}:
\begin{equation} \label{eq:pe-threshold-n1}
    \pe^* =  \left ( \frac{L}{\ell_h
    }\right )^2  \frac{\left(1 + k_1^2 \ell_h^2 \right)^2}{c_0 f'(c_0)\left(1 + k_1^2\ell_h^2 + \frac{1}{2}\eps^2 \right)} \,,
\end{equation}
which decreases with increasing friction inhomogeneity magnitude $\eps$, and agrees for $\eps=0$ with Ref.~\cite{boisPatternFormationActive2011}. Figure~\ref{fig:dispersion_relation_and_stability}b shows the corresponding stability diagram for a relative hydrodynamic length $\ell_h/L=0.2$. The approximate result Eq.~(\ref{eq:pe-threshold-n1}) (dashed line) only deviates from the numerically obtained exact transition prediction (solid line) for larger $\eps$. Numerical simulations of the fully non-linear system confirm the exact transition prediction. The patterns that form when $\pe>\pe^*$ (red dots) are similar to those in Fig.~\ref{fig:time_series_and_v_init_diagram}a. 
The monotonic decay of $\pe^*$ shown in Fig.~\ref{fig:dispersion_relation_and_stability}b implies that increasing friction inhomogeneity (at fixed \pec{} number $\pe$) can itself cause mechanochemical pattern formation. We find that hydrodynamic screening is necessary for this to occur. Figure~\ref{fig:dispersion_relation_and_stability}c shows critical P\'eclet numbers $\pe^*$ as function of $\eps$ for different hydrodynamic screening lengths $\ell_h$, where solid (dashed) lines show numerically exact (approximate) predictions. When the hydrodynamic screening length approaches the system size ($\ell_h/L\sim1$) the critical \pec{} number $\pe^*$ becomes essentially independent of the friction inhomogeneity magnitude~$\eps$. This shows that systems with long-ranged active flows -- facilitated by large hydrodynamic screening lengths~$\ell_h$ -- are less sensitive to friction that varies over smaller length scales. In contrast, smaller hydrodynamic screening lengths $\ell_h<L$ facilitate the nucleation of active flow instabilities near local friction minima where $\gamma(x)<\gamma_0$, which leads to a reduced critical \pec{} number~$\pe^*$ compared to the homogeneous case where $\gamma(x)=\gamma_0$.

\begin{figure}
    \centering
    \includegraphics{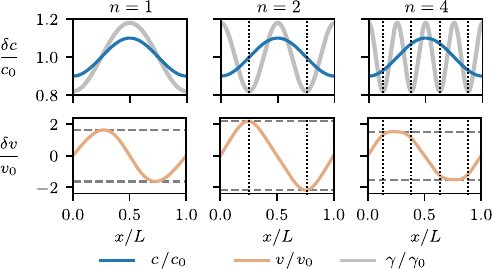}
    \caption{Unstable concentration modes (blue lines, top) which typically arise in the absence of degradation ($r=0$) lead to flows (orange lines, bottom) whose extrema optimally align with friction minima (vertical dashed lines) when $n=2$ (see Eq.~\eqref{eq:fric-function}). This leads to non-monotonic behavior of critical \pec{} number for decreasing friction pattern length scale $\ell_f\sim L/n$ (see Fig.~\ref{fig:dispersion_relation_and_stability}d). }
    \label{fig:v_affected_by_fric_wavelength}
\end{figure}

To investigate the connection between hydrodynamic screening and friction inhomogeneity further, we fix $\ell_h/L=0.2$ and vary the length scale of friction variations $\ell_f\sim L/n$ by changing $n$ in Eq.~(\ref{eq:fric-function}) (Fig.~\ref{fig:dispersion_relation_and_stability}d). Now, for most values of \eps{}, the critical P\'eclet number~$\pe^*$ exhibits unexpected non-monotonic behavior, where an intermediate friction length scale ($n=2$) minimizes the critical activity required for mechanochemical instabilities. This can be understood from the spatial profile of the flow induced by the dominant component \mbox{$\delta c=\delta c_1 e^{i 2\pi x/L}$} of the unstable concentration eigenmode (Fig.~\ref{fig:v_affected_by_fric_wavelength}, top row). By symmetry, the associated flow profile~$\delta v$ (Fig.~\ref{fig:v_affected_by_fric_wavelength}, bottom row) must contain two roughly evenly spaced extrema. These align almost perfectly with the two minima in the friction pattern Eq.~(\ref{eq:fric-function}) with $n=2$, but always have to be at least partially in regions of higher friction when $n\ne2$. Consequently, the lowest activity, i.e. the smallest \pec{} number, is needed for an instability when $n=2$.

\label{sec:results:subsec:friction_guided_pattern_formation}
\begin{figure*}
    \includegraphics{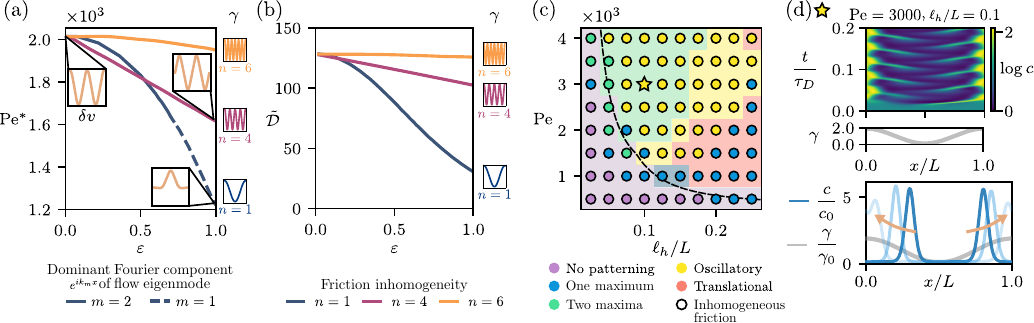}
    \caption{Impact of inhomogeneous friction on active fluids with intrinsic pattern length scale selection. (a)~Critical \pec{} numbers for different friction pattern length scales $\ell_f=L/n$ ($n\in\{1,4,6\}$) when modes with smaller wavelength (here $m=2$, see inset of flow eigenmodes $\delta v$ at $\eps=0$) become unstable first in the homogeneous system. Dashed line indicates regime where the $m=2$ selection is suppressed in favor of a mode with dominant $m=1$ component (see inset $n=1$, $\eps=1$). $\ell_h/L=0.1, r=150$ (b)~Total dissipated power $\tilde{\mathcal{D}}$ [see Eq.~(\ref{eq:dissint})] for parameters from (a) decays most rapidly for friction pattern with $n=1$. (c)~Phase diagram comparing dynamics in homogeneous ($\eps=0$, background shading) and inhomogeneous ($\eps=0.9$, colored circles) systems. Dashed line indicates exact stability boundary for $\eps=0$ [$\lambda=0$ in Eq.~(\ref{eq:homog-dispersion-relation})]. Inhomogeneous friction arrests translational dynamics (blue circles on red shading) and leads to oscillations when steady state and friction patterns become incommensurable (yellow circles on green shading). $n=1$, $r=50$ (d)~Kymograph (top) of oscillatory patterning dynamics in which double-peaks (steady-state of the homogeneous system) repeatably form, climb up friction gradients, and collide (bottom).}
    \label{fig:degredation_and_higher_order}
\end{figure*}

\subsection{Active fluid with intrinsic length scale selection}\label{sec:results:subsec:higher_order}    

We now discuss the role of inhomogeneous friction in systems with degradation (\mbox{$r>0$}), a scenario allowing modes with smaller intrinsic wavelengths to become unstable first. Specifically, we consider parameters for which -- in a system with homogeneous friction ($\varepsilon=0$) -- modes \mbox{$\delta c=\delta c_m e^{i 2\pi m x/L}$} with $|m|=2$ (as opposed to $|m|=1$ discussed before) is the first to become unstable above some critical~\pec{} number (gray lines in Fig.~\ref{fig:dispersion_relation_and_stability}a). Introducing inhomogeneous friction ($\eps>0$) changes~$\pe^*$ as shown in Fig.~\ref{fig:degredation_and_higher_order}a. Similar to the previous scenario (see Fig.~\ref{fig:dispersion_relation_and_stability}d), the critical P\'eclet number~$\pe^*$ varies non-monotonically when decreasing the friction inhomogeneity length scale $\ell_f$, where now $g(x)$ [see Eq.~(\ref{eq:fric-function})] with $n=4$ yields minimal values for~$\pe^*$ (Fig.~\ref{fig:degredation_and_higher_order}a, $\eps\lesssim0.5$). This can again be understood from the commensurability between the flows triggered by the instability and the specific friction pattern: the dominant mode at the instability now has four extrema (Fig.~\ref{fig:degredation_and_higher_order}a, left inset) and hence best aligns with the four minima of the friction profile Eq.~(\ref{eq:fric-function}) with~$n=4$.

For larger friction inhomogeneity ($\eps\gtrsim0.5$), we find a crossover above which the lowest critical \pec{} number is associated with the longest friction length scale $\ell_f\sim L$ (Fig.~\ref{fig:degredation_and_higher_order}a, $n=1$). In this regime, strong mode coupling generated by the friction inhomogeneity overrides the shorter wavelength instability preferred by the homogeneous system (Fig.~\ref{fig:degredation_and_higher_order}a, dashed line). Specifically, the dominant unstable eigenmode leads to flow profiles with one prominent maximum (Fig.~\ref{fig:degredation_and_higher_order}a, right inset), which experience the least resistance from a friction pattern Eq.~(\ref{eq:fric-function}) with $n=1$. 

To understand this crossover, and inspired by the important role of flow and friction length scales above, we consider the total power dissipated by flow eigenmodes
\begin{equation}\label{eq:dissint}
    \mathcal{D    }[\delta v] = \frac{1}{2} \int\mathrm{d}x\left[\eta(\partial_x \delta v)^2 + \gamma_0g(x) (\delta v)^2\right].
\end{equation}
In Fig.~\ref{fig:degredation_and_higher_order}b we plot the non-dimensional dissipation, $\tilde{\mathcal{D}}/\mathcal{D}_0$ with $\mathcal{D}_0=L\eta/\tau_D^2$. This shows that, while $\tilde{\mathcal{D}}[\delta v]$ decreases for all flow eigenmodes $\delta v$ as $\eps$ increases (consistent with the decreasing critical \pec{} numbers seen in Fig.~\ref{fig:degredation_and_higher_order}a), it does so most rapidly for modes varying at system-sized length scales $\delta v\approx\text{Re}\left(e^{i2\pi x/L}\right)$. This mode experiences the least dissipation for a friction inhomogeneity profile Eq.~(\ref{eq:fric-function}) with $n=1$, which ultimately leads to the crossover in the critical \pec{} number observed in~Fig.~\ref{fig:degredation_and_higher_order}a.

\subsection{Mechanochemical frustration and oscillatory patterns}\label{sec:non-linear-states}

The previous discussion has focused on how inhomogeneous friction causes mode coupling and alters critical transition parameters in a self-organized active fluid. Finally, we use these insights to rationalize the non-linear response of this system in the presence of inhomogeneous friction. To this end, we fix a friction pattern Eq.~(\ref{eq:fric-function}) with $n=1$ and inhomogeneity magnitude $\eps=0.9$, classify the non-linear steady-state patterns emerging from Eqs~(\ref{eq:conservation}) and (\ref{eq:force-balance}) over a large range of \pec{} numbers and hydrodynamic length scales (colored circles in Fig.~\ref{fig:degredation_and_higher_order}c), and compare the results with the corresponding homogeneous scenarios ($\eps=0$, background shading in Fig.~\ref{fig:degredation_and_higher_order}c). The region of a stable homogeneous state for $\eps=0$ (purple shading) is accurately predicted by the linear stability analysis (Eq.~(\ref{eq:homog-dispersion-relation}) with $\lambda=0$, black dashed line). 

With inhomogeneous friction, steady states with one or two maxima (blue and green circles, respectively) appear at lower \pec{} numbers, consistent with the analysis of critical \pec{} numbers in Figs.~\ref{fig:dispersion_relation_and_stability}c and~\ref{fig:dispersion_relation_and_stability}d. In parts of the unstable regime, the homogeneous system exhibits states of translating concentration patterns (Fig.~\ref{fig:degredation_and_higher_order}c, red shading). Inhomogeneous friction breaks translational invariance and largely arrests this translation, leading to stationary patterns (blue circles). Furthermore, our analysis reveals a parameter regime of mechanochemical frustration. For smaller hydrodynamic lengths $\ell_h$, where the homogeneous system spontaneously forms stable patterns with two concentration maxima (Fig.~\ref{fig:degredation_and_higher_order}c, green shading), inhomogeneous friction leads to oscillatory dynamics (yellow circles). Figure~\ref{fig:degredation_and_higher_order}d shows an exemplary kymograph of this dynamics (parameters indicated with yellow star in Fig.~\ref{fig:degredation_and_higher_order}c). The emergence of such oscillations can be understood from the incommensurability of the pattern favored by the homogeneous system for these parameters (two concentration maxima) and the symmetry of the friction pattern with only one maximum (Eq.~(\ref{eq:fric-function}) with $n=1$). Pairs of concentration maxima keep forming but have to move up friction gradients in opposite directions and eventually collide, giving time and space for new peaks to form in the low-concentration region that emerges from that process.

\section{Conclusion}\label{sec:conclusion}
We have introduced a minimal model of a self-organized active fluid that experiences inhomogeneous mechanical interactions with the surroundings. We have shown that contractile patches move up friction gradients, identified an intricate interplay of different length scales associated with hydrodynamic screening, friction inhomogeneities and diffusion-degradation that sets critical \pec{} numbers, and found oscillatory states that emerge from a frustration between fluid-intrinsic pattern selection and external friction patterns. 

The phenomenology described in this work is most likely to be observed in systems where hydrodynamic screening is relevant ($\ell_h/L<1$), which we would expect to be the case for active materials that strongly adhere -- or are in close contact with -- a substrate, egg shell or membrane. This includes cells that migrate on substrates or evolve in tight confinement, as well as reconstituted actomyosin networks that adhere to substrates or membranes \cite{abushahSymmetryBreakingReconstituted2014,pimpaleCellLineagedependentChiral2020,thor11,carvalhoCellsizedLiposomesReveal2013}. Measuring a hydrodynamic screening length directly is challenging, but instances of cortical flows and tissue motion were found to be consistent with $\ell_h/L~\sim~0.15-0.7$~\cite{mayerAnisotropiesCorticalTension2010,pimpaleCellLineagedependentChiral2020,pfanzelterActiveTorqueDipole2025}. Exciting oscillations of active stress patterns by introducing external inhomogeneities would additionally require large \pec{} numbers ($\pe{} \gg 1$), i.e. high levels of contractility or weak diffusion. An estimate based on \textit{in vivo }cortical flow speeds around 5\, m/min over a length of~10~\textmu m~\cite{reymann_cortical_2016}, and a actomyosin diffusivity of~$10^{-3}$\textmu m$^2$/s~\cite{weirichLiquidBehaviorCrosslinked2017, luoAnalysisLocalOrganization2013} yields $\pe \approx80$. While this estimate falls short of the values suggested in Fig.~4, it does not explicitly account for difficult-to-measure contractility values and may therefore still be missing prefactors for a fully quantitative comparison with the \pec{} number identified in our model. Alternatively, one might consider synthetic systems based on reconstituted actin networks and exploit subdiffusive transport properties~\cite{wong04,brunoTransitionSuperdiffusiveBehavior2009} or tunable contractility~\cite{saka24} to obtain a system with a larger effective \pec{} number.

Our work can naturally be extended to study more complex patterns of external friction, as well as to investigate the impact of multiple regulator species and a mechano-sensitive regulator dynamics, both of which lead to richer patterning behavior even when friction is homogeneous~\cite{kumarPulsatoryPatternsActive2014,bonatiRoleMechanosensitiveBinding2022,leberreGeometricFrictionDirects2013,shellardFrictiotaxisUnderliesFocal2025}. Such extensions should shed further light on general principles of how spontaneous pattern formation in active materials can be guided by externally-broken symmetries.

%\bibliography{active_fluids_refs_abbrev}

%

\clearpage
\onecolumngrid
\renewcommand\thefigure{S\arabic{figure}}    
\setcounter{figure}{0} 
\appendix

\section{Nondimensional equations} \label{sec:appendix}
\setcounter{page}{1}

We non-dimensionalize coordinates $x$, time $t$, velocity $v$ and concentration $c$ by the domain length $L$, the diffusion timescale $\tau_D=L^2/D$, the characteristic velocity $D/L$, and a reference concentration $c_0$, respectively. We represent these non-dimensionalised quantities with tildes. We also introduce a characteristic hydrodynamic lengthscale $\ell_h = \sqrt{\eta/\gamma_0}$ \cite{mayerAnisotropiesCorticalTension2010}, and the non-dimensional \pec{} number, $\pe = \xi L^2/\eta D$, describing the ratio of the diffusive and advective timescales.

With this, force balance (Eq.~(\ref{eq:force-balance}), main text) and continuity equations (Eq.~(\ref{eq:conservation}), main text) can be written as
\begin{align}
    \partial_{\tilde{x}}^{2}\tilde{v} - \alpha^{-1} g(\tilde{x})\tilde{v} + \text{Pe} \partial_{\tilde{x}} f({c}) = 0 \label{eq:FBnd}\\
    \partial_{\tilde{t}} \tilde{c} = \partial^2_{\tilde{x}} \tilde{c} - \partial_{\tilde{x}}(\tilde{v} \tilde{c}) -\tilde{r}(\tilde{c}-1),\label{eq:CEnd}
\end{align}
where $\tilde{r}=r\tau_D$ is the dimenionsless degradation rate and
\begin{equation}
\alpha=\frac{\ell_h^2}{L^2}   
\end{equation}
is the dimensionless hydrodynamic length (squared). Wavenumbers of the Fourier modes are accordingly $\tilde{k}_m=Lk_m=2\pi m$. For brevity, we omit the tildes again in the equations to follow in this supplement.

\section{Linear stability analysis}
\subsection{Linearization}

We linearize Eqs.~(\ref{eq:FBnd}) and (\ref{eq:CEnd}) around the homogeneous state, \mbox{$c = 1 + \delta c, v = \delta v$} to obtain:
\begin{align}
    \partial_x^2 \delta v - \alpha^{-1} g(x) \delta v + \pe f'(1) \partial_x \delta c &= 0 \label{eq:FBnd}\\
    \partial_t \delta c = \partial_x^2 \delta c - \partial_x \delta v - r\delta c
\end{align}

We then expand $\delta c$, $\delta v$, as well as the friction pattern $g(x)$ in Fourier modes as
\begin{align}
    \delta c &= \sum_{m\in\mathbb{Z}} \delta c_m e^{\lambda_m t + i k_m x}\label{eq:fec}\\
    \delta v &= \sum_{m\in\mathbb{Z}} \delta v_m e^{\lambda_m t + i k_m x} \\
    g(x) &= \sum_{m\in\mathbb{Z}} g_m e^{i k_m x}\label{eq:feg},
\end{align}
where $k_m=2\pi m$. A non-constant friction pattern leads to mode coupling via the term
\begin{equation}
    g(x) \delta v = \sum_{m,m'\in\mathbb{Z}}g_{m}\delta v_{m'} e^{i (k_m + k_{m'}) x},
\end{equation}
and thereby alters instability thresholds. Using the expansions Eqs.~(\ref{eq:fec})-(\ref{eq:feg}) in the force-balance Eq.~(\ref{eq:FBnd}) and isolating a single mode yields
\begin{equation}\label{eq:meFB}
    k_l^2 \delta v_l + \alpha^{-1} (g * \delta v)_l = i f'(1) \pe \, k_l \delta c_l,
\end{equation}
where $(g * \delta v)_l = \sum_m{g_{l-m} \delta v_{m}}$ denotes a discrete convolution. We can express the left-hand side of Eq.~(\ref{eq:meFB}) as
\begin{equation}
    [\mmat \delta v]_l = k_l^2 \delta v_l + \alpha^{-1}\sum_{m\in\mathbb{Z}} g_{l-m} \delta v_{m}, 
\end{equation}
where the operator $\mmat$ is given by
\begin{equation}
    \mmat_{lm} = k_m^2 \delta _{lm} + \alpha^{-1}g_{l-m}.
\end{equation}
Finding the velocity mode coefficients $\delta v_l$ therefore amounts to determining
\begin{equation}\label{eq:dvm}
    \delta v_l = i f'(1) \pe \sum_{m\in\mathbb{Z}} [\mmat^{-1}]_{lm} k_m \delta c_m.
\end{equation}

Using the expansion Eq.~(\ref{eq:fec}) in the continuity equation~(\ref{eq:CEnd}) and eliminating velocity mode coefficients using Eq.~(\ref{eq:dvm}) yields the eigenvalue problem
\begin{align} \label{eq:app:a-matrix-homogeneous}
\lambda \delta c_l = \left[\mathcal{J} \delta c\right]_l, 
\end{align}
where $\mathcal{J}$ is the Jacobian with components
\begin{equation}\label{eq:jac}
\mathcal{J}_{lm} = - (k_l^2 + r) \delta_{lm} + \pe f'(1) k_l k_m [\mmat^{-1}]_{lm}.
\end{equation}

For homogeneous friction, $\eps=0$, we have $g_k=1$ for $k=0$, otherwise $g_k=0$ (equivalently $g_{l-m}=\delta_{lm}$), such that matrix $\mmat_{lm}\sim\delta_{lm}$ and the dispersion determined by Eq.~(\ref{eq:app:a-matrix-homogeneous}) reads
\begin{equation} \label{eq:dispersion-relation-bois}
    \lambda^{(0)}_m = k_m^2 \left(\frac{\alpha }{1 + k_m^2\alpha}f'(1)\pe  - 1 \right)-r,
\end{equation}
which is in the context of this work the zeroth order contribution to the growth rate, equivalent to the result found in Ref.~\cite{boisPatternFormationActive2011}.

\subsection{Structure of \mmat} \label{sec:app:M}
Consider for simplicity a friction pattern expanded as in Eq.~(\ref{eq:feg}) made of a single non-trivial mode in addition to $g_0$. The matrix $\mmat$ is in this case tridiagonal. The inverse of such matrix is non-zero at the same three diagonals as \mmat\ itself, as well as repeatedly at all further equally spaced off-diagonals~\cite{smolarskiDiagonallystripedMatricesApproximate2006}. For example, if the components $\mmat_{lm}$ with $|l-m|=2$ contain a non-zero values, then the inverse will have non-zero elements on all off-diagonals with $|l-m|=2s$ ($s\in\mathbb{Z}$).\\

In our case, where $g(x)$ is of the form $1 + \eps \cos (2 \pi n x)$, the matrix $\alpha \mmat$ is tridiagonal with $1 + k_m^2 \alpha$ along the main diagonal and $\eps/2$ along the $\pm n$'s off-diagonals. Explicitly, for different $n$ it takes the form (components $\mmat{}_{lm}$ truncated here at $|l|,|m|=2$ and $\mmat{}_{00}$ placed at the center)
%\def\0{\ensuremath{\mathcolor{lightgray} 0}}
%\begin{equation}
%\label{eq:M1}
%\alpha\mmat^{n=1}=
%\begin{tikzpicture}[baseline=(m.center)]
%\matrix (m) [matrix of math nodes,
%             left delimiter={(}, right delimiter={)},
%             row sep=-2pt, column sep=0pt] {
%1+k_{-2}^2 \alpha & \frac{\eps}{2} & \0 & \0 & \0 \\
% \frac{\eps}{2} & 1+k_{-1}^2 \alpha & \frac{\eps}{2} & \0 & \0 \\
%\0 & \frac{\eps}{2} & \hphantom{kk} 1 \hphantom{kk}  & \frac{\eps}{2} & \0 \\
% \0 & \0 & \frac{\eps}{2} & 1 + k_1^2 \alpha & \frac{\eps}{2} \\
% \0 & \0  & \0 & \frac{\eps}{2} & 1 + k_2^2 \alpha  \\
%};
%\node[draw,dashed,rounded corners,inner xsep=18pt,inner ysep=1pt,fit=(m-4-2) (m-2-4)] {};
%\end{tikzpicture},
%\end{equation}

\begin{equation}
\label{eq:M1}
\alpha\mmat^{n=1}=
\begin{tikzpicture}[baseline=(m.center)]
\matrix (m) [matrix of math nodes,
             left delimiter={(}, right delimiter={)},
             row sep=-2pt, column sep=0pt] {
1+k_{-2}^2 \alpha & \frac{\eps}{2} & \grayzero & \grayzero & \grayzero \\
\frac{\eps}{2} & 1+k_{-1}^2 \alpha & \frac{\eps}{2} & \grayzero & \grayzero \\
\grayzero & \frac{\eps}{2} & \hphantom{kk} 1 \hphantom{kk} & \frac{\eps}{2} & \grayzero \\
\grayzero & \grayzero & \frac{\eps}{2} & 1 + k_1^2 \alpha & \frac{\eps}{2} \\
\grayzero & \grayzero & \grayzero & \frac{\eps}{2} & 1 + k_2^2 \alpha \\
};
\node[draw,dashed,rounded corners,inner xsep=18pt,inner ysep=1pt,
      fit=(m-4-2) (m-2-4)] {};
\end{tikzpicture},
\end{equation}

%\begin{equation}
%\label{eq:M2}
%\alpha\mmat^{n=2}=
%\begin{tikzpicture}[baseline=(m.center)]
%\matrix (m) [matrix of math nodes,
%             left delimiter={(}, right delimiter={)},
%             row sep=-2pt, column sep=0pt] {
%1+k_{-2}^2 \alpha & \0 & \frac{\eps}{2} & \0 & \0 \\
% \0 & 1+k_{-1}^2 \alpha & \0 & \frac{\eps}{2} & \0 \\
%\frac{\eps}{2} & \0 & \hphantom{kk} 1 \hphantom{kk}  & \0 & \frac{\eps}{2} \\
% \0 & \frac{\eps}{2} & \0 & 1 + k_1^2 \alpha & \0 \\
% \0 & \0  & \frac{\eps}{2} & \0 & 1 + k_2^2 \alpha \\
%};
%\node[draw,dashed,rounded corners,inner xsep=18pt,inner ysep=1pt,fit=(m-4-2) (m-2-4)] {};
%\end{tikzpicture},
%\end{equation}

\begin{equation}
\label{eq:M2}
\alpha\mmat^{n=2}=
\begin{tikzpicture}[baseline=(m.center)]
\matrix (m) [matrix of math nodes,
             left delimiter={(}, right delimiter={)},
             row sep=-2pt, column sep=0pt] {
1+k_{-2}^2 \alpha & \grayzero & \frac{\eps}{2} & \grayzero & \grayzero \\
\grayzero & 1+k_{-1}^2 \alpha & \grayzero & \frac{\eps}{2} & \grayzero \\
\frac{\eps}{2} & \grayzero & \hphantom{kk} 1 \hphantom{kk} & \grayzero & \frac{\eps}{2} \\
\grayzero & \frac{\eps}{2} & \grayzero & 1 + k_1^2 \alpha & \grayzero \\
\grayzero & \grayzero & \frac{\eps}{2} & \grayzero & 1 + k_2^2 \alpha \\
};
\node[draw,dashed,rounded corners,inner xsep=18pt,inner ysep=1pt,
      fit=(m-4-2) (m-2-4)] {};
\end{tikzpicture},
\end{equation}

and analog for increasing $n$. Infinite matrices of this form do in general not have closed expressions for the components of their inverses. For the analytic analysis, we therefore consider truncated matrices \mmat{} like the ones shown in Eqs.~(\ref{eq:M1}), (\ref{eq:M2}). To study the impact of variable friction as quantified by $\eps$, a minimal requirement for this truncation is to maintain sufficiently many off-diagonals so that $\eps$ still contributes to \mmat. For both the matrices shown in Eqs.~(\ref{eq:M1}), (\ref{eq:M2}), this amounts to a truncation at $|l|,|m|=1$ (sub-matrices enclosed by dashed boxes).

\subsection{Spectrum of \amat{}} \label{app:sec:spectrum}
With homogeneous friction, the Jacobian \amat{} defined in Eq.~(\ref{eq:jac}) is diagonal and its eigenvalues come in degenerate pairs, i.e. $\lambda^{(0)}(k_m) = \lambda^{(0)}(k_{-m})$ [see Eq.~\eqref{eq:dispersion-relation-bois}]. The general Jacobian \amat{} with $\eps\ne0$ is still symmetric, such that its eigenvalues are still real and an orthogonal basis of eigenvectors exists. However, these eigenvectors will in general no longer be pure modes of the form $e^{ik_mx}$. To systematically approximate how inhomogeneous friction changes the spectrum of the Jacobian, we therefore have to analyze \amat{} using spectral perturbation theory~\cite{landauQuantumMechanics1977}.

\subsubsection{Growth rate corrections of $|m|=1$ modes} 
To illustrate this and derive Eq.~(\ref{eq:pe-threshold-n1}) (main text), consider first the case $r=0$ and \mbox{$g(x)= 1 + \eps \cos(2 \pi x)$}. The minimally truncated matrix \mmat{} is then given by

\begin{align}
        \alpha\mmat^{n=1}&=\begin{pmatrix}
        {1 + k_1^2 \alpha} & \frac{\eps}{2} & 0 \\
        \frac{\eps}{2} & 1 & \frac{\eps}{2} \\
        0 & \frac{\eps}{2} & {1+k_1^2\alpha}
    \end{pmatrix}.
\end{align}

Denoting $\beta_1=1/(1+k_1^2\alpha)>1$, the inverse of this matrix has an expansion in $\eps$ given by
\begin{equation}
\alpha^{-1}\left(\mmat^{n=1}\right)^{-1}= 
    \begin{pmatrix}
        \beta_1 & 0 & 0 \\
        0 & 1 & 0 \\
        0 & 0 & \beta_1
    \end{pmatrix}
    + \frac{\beta_1}{2}
    \begin{pmatrix}
        0 & -1 & 0 \\
        -1 & 1 & -1 \\
        0 & -1 & 0
    \end{pmatrix}\varepsilon + 
    \frac{\beta_1^2}{4}
     \begin{pmatrix}
        1 & 0 & 1 \\
        0 & 2 & 0 \\
        1 & 0 & 1
    \end{pmatrix} \varepsilon^2 + \mathcal{O}(\eps^3),
    \label{eq:app:n1-expansion-example}
\end{equation}
corresponding to a decomposition into a diagonal part and perturbations in increasing powers of \eps{}. The non-vanishing matrix elements of the $\mathcal{O}(\eps)$ term are multiplied by $k_0=0$ in the Jacobian Eq.~(\ref{eq:jac}) and do therefore not contribute. The block form of the~$\mathcal{O}(\eps^2)$ term in Eq.~(\ref{eq:app:n1-expansion-example}) decouples in the Jacobian the response of the $m=0$ mode -- marginal (stable) for $r=0$ ($r>0$) -- and the combined response of a vector containing contributions from the Fourier modes $m=1$ and $m=-1$. Using Eq.~(\ref{eq:app:n1-expansion-example}) in Eq.~(\ref{eq:jac}) yields an effective Jacobian for the block of components $(-1,-1), (-1,1), (1,-1)$ and $(1,1)$ that can be written~as
\begin{align}\label{eq:jachat}
\hat{\mathcal{J}}_1 &=\hat{\mathcal{J}}^{(0)}_{1}+\eps^2\hat{\mathcal{J}}^{(1)}_{1}\\
&:=-\left[r+k_1^2(1-\alpha\beta_1\pe\,f'(1))\right]\mathbb{I}
+\eps^2k_1^2\frac{\alpha\beta_1^2}{4}\pe\,f'(1) \label{eq:jachatfull}
\begin{pmatrix}
1 & -1\\
-1 & 1
\end{pmatrix}.
\end{align}

The eigenvectors of $\hat{\amat}^{(1)}_1$ are $(1,1)^T$ and $(1,-1)^T$, which trivially also form an eigenbasis of $\hat{\amat}^{(0)}_1$. However, while the eigenvalues of $\hat{\amat}^{(0)}_1$ -- given in Eq.~(\ref{eq:dispersion-relation-bois}) -- are degenerate, they are distinct for $\hat{\amat}^{(1)}_1$, which has eigenvalues
\begin{align}
\lambda^{(1)}_{1,a}=0\qquad \lambda^{(1)}_{1,b}=k_1^2\frac{\alpha\beta_1^2}{2}\pe\,f'(1).   
\end{align}
We therefore find the effective Jacobian Eq.~(\ref{eq:jachatfull}) has -- up to first order in perturbation theory -- eigenvalues
\begin{align} \label{eq:app:k1_n1_eigenvalue_final_corrections}
% \hat{\lambda}_{1,a} &= \lambda^{(0)}_1+\eps^2\lambda^{(1)}_{1,a}=-r - k_1^2 + \frac{\pe\,f'(1) k_1^2\alpha}{1 + k_1^2 \alpha}\\    
% \hat{\lambda}_{1,b} &=\lambda^{(0)}_1+\eps^2\lambda^{(1)}_{1,b}= -r - k_1^2 + \pe\,f'(1) k_1^2\alpha \left [ \frac{1}{1 + k_1^2 \alpha} + \frac{2}{(1 + k_1^2 \alpha)^2} \left (\frac{\eps}{2}\right )^2 \right ]\label{eq:app:disp_rel_corrected_k1_n1},
\hat{\lambda}_{1,a} &= \lambda^{(0)}_1+\eps^2\lambda^{(1)}_{1,a}=-r - k_1^2 + \pe\,f'(1) k_1^2\alpha \beta_1 \\    
\hat{\lambda}_{1,b} &=\lambda^{(0)}_1+\eps^2\lambda^{(1)}_{1,b}= -r - k_1^2 + \pe\,f'(1) k_1^2\alpha \left [ \beta_1 + 2 \beta_1^2 \left (\frac{\eps}{2}\right )^2 \right ]\label{eq:app:disp_rel_corrected_k1_n1},
\end{align}
which corresponds to a continuous eigenvalue splitting when $\varepsilon>0$ that lifts the degeneracy present at $\varepsilon=0$. This loss of symmetry in the eigenvalues is associated with the loss of continuous translational symmetry of the equations of motion when~$\eps\ne0$. Equation~(\ref{eq:app:disp_rel_corrected_k1_n1}) is the approximation shown in the main text in Fig.~\ref{fig:dispersion_relation_and_stability}a (black dashed lines). Setting $\hat{\lambda}_{1,b}=0$ and rearranging \eqref{eq:app:disp_rel_corrected_k1_n1} yields the critical \pec{} number approximations Eq.~(\ref{eq:pe-threshold-n1}) that are shown as gray dashed lines in Fig.~\ref{fig:dispersion_relation_and_stability}b,c.\\

The splitting of the eigenvalues is illustrated in Fig.~\ref{fig:app:spectrum}a, where the $|m|=1$ eigenvalues are degenerate when $\eps=0$, and split according to Eqs.~\eqref{eq:app:k1_n1_eigenvalue_final_corrections},(\ref{eq:app:disp_rel_corrected_k1_n1}) when $\eps>0$. Solid lines show for homogeneous friction ($\eps=0$) the eigenvalues $\lambda_m^{(0)}$ for $m=0$ [see Eq.~(\ref{eq:dispersion-relation-bois})] and $m=1$ [see Eqs.~\eqref{eq:app:k1_n1_eigenvalue_final_corrections} and (\ref{eq:app:disp_rel_corrected_k1_n1})]. Dashed blue lines depict approximate results, blue points show the numerically exact eigenvalues of the full Jacobian. The change of sign in the growth rate $\hat{\lambda}_{1,b}$ indicates that friction inhomogeneities can promotes the spontaneous formation of mechanochemical patterns.

\subsubsection{Growth rate corrections of $|m|=2$ modes} 

As discussed in Section \ref{sec:results:subsec:higher_order}, a degradation rate $r>0$ can cause higher-order Fourier modes (such as those with $|m|=2$) to become unstable first. To analyze the effect of inhomogeneous friction on the growth rates of these modes, we must consider a 5$\times$5 similar to Eq.~\eqref{eq:M1}. For the same friction pattern $g(x)$ as in the previous section, the minimally truncated matrix \mmat{} is:
\begin{equation}
\alpha \mmat{}^{n=1} = 
\begin{pmatrix}
    1+k_{-2}^2 \alpha & \frac{\eps}{2} & 0 & 0 & 0 \\
 \frac{\eps}{2} & 1+k_{-1}^2 \alpha & \frac{\eps}{2} & 0 & 0 \\
0 & \frac{\eps}{2} & \hphantom{kk} 1 \hphantom{kk}  & \frac{\eps}{2} & 0 \\
 0 & 0 & \frac{\eps}{2} & 1 + k_1^2 \alpha & \frac{\eps}{2} \\
 0 & 0 & 0  & \frac{\eps}{2} & 1 + k_2^2 \alpha \\
 \end{pmatrix}.
\end{equation}

Denoting $\beta_1 = 1/(1 + k_1^2 \alpha)$ and $\beta_2 = 1/(1+k_2^2 \alpha)$, we find
\begin{align}
\alpha^{-1} (\mmat{}^{n=1})^{-1} &= 
\begin{pmatrix}
\beta_2 & 0 & 0 & 0 & 0 \\
0 & \beta_1 & 0 & 0 & 0 \\
0 & 0 & 1 & 0 & 0 \\
0 & 0 & 0 & \beta_1 & 0 \\
 0 & 0 & 0 & 0 & \beta_2 \\
\end{pmatrix}
+
\begin{pmatrix}
0 & - \frac{\beta_1 \beta_2}{2} & 0 & 0 & 0 \\
- \frac{\beta_1 \beta_2}{2} & 0 & -\frac{\beta_1}{2} & 0 & 0 \\
0 & - \frac{\beta_1}{2} & 0 & -\frac{\beta_1}{2} & 0 \\
0 & 0 & -\frac{\beta_1}{2} & 0 & -\frac{\beta_1 \beta_2}{2} \\
0 & 0 & 0 & -\frac{\beta_1 \beta_2}{2} & 0 \\
\end{pmatrix} \eps \notag\\
&+
\begin{pmatrix}
\frac{\beta_1 \beta_2^2}{4} & 0 & -\frac{\beta_1}{12} + \frac{\beta_2}{3} & 0 & 0 \\
0 & -\frac{\beta_1}{9} + \frac{\beta_1^2}{6} + \frac{4\beta_2}{9} & 0 & \frac{\beta_1^2}{4} & 0 \\
 -\frac{\beta_1}{12} + \frac{\beta_2}{3} & 0 & \frac{\beta_1}{2} & 0 & -\frac{\beta_1}{12} + \frac{\beta_2}{3}\\
0 & \frac{\beta_1^2}{4} & 0 & -\frac{\beta_1}{9} + \frac{\beta_1^2}{6} + \frac{4\beta_2}{9} & 0 \\
0 & 0 & -\frac{\beta_1}{12} + \frac{\beta_2}{3} & 0 & \frac{\beta_1 \beta_2^2}{4} \\
\end{pmatrix} \eps^2
+ \mathcal{O}(\eps^3).\label{eq:Mm2}
\end{align}

Using the matrix Eq.~(\ref{eq:Mm2}) in Eq.~(\ref{eq:jac}), we find that the $\mathcal{O}(\eps)$ term has vanishing components at $(\pm2, \pm2)$ and therefore does not contribute to spectrum in the (degenerate) subspace of interest spanned by $|-2\rangle:=(1,0,0,0,0)$ and $|2\rangle:=(0,0,0,0,1)$. The $\mathcal{O}(\eps^2)$ term in Eq.~(\ref{eq:Mm2}) on the other does contribute diagonal couplings in that subspace, such that perturbation theory yields an effective Jacobian of the of the form
\begin{align}
    \hat{\mathcal{J}}_2 &= - \left [ r + k_2^2(1 - \alpha \beta_2 \pe f'(1) ) \right ] \mathbb{I} + \eps^2 k_2^2 \alpha \pe f'(1) 
    \left ( 
    \frac{\beta_1 \beta_2^2}{4} + \frac{\beta_1^2 \beta_2^2}{4(\beta_2-\beta_1)}
    \right )
    \begin{pmatrix}
    1 & 0\\
    0 & 1
    \end{pmatrix},
\end{align}
where the second term is a consequence of the off-diagonal couplings in the $\mathcal{O}(\eps^2)$ matrix in Eq.~(\ref{eq:Mm2}). The eigenvectors in this case are trivial, and the eigenvalues are given by
\begin{equation}
\hat{\lambda}_2^{(1)} = \lambda_2^{(0)} + \eps^2 \lambda_2^{(1)} := - r - k_2^2 + \pe f'(1)k_2^2 \alpha \left [ 
\beta_2 + 
\left ( 
    \frac{\beta_1 \beta_2^2}{4} + \frac{\beta_1^2 \beta_2^2}{4(\beta_2-\beta_1)}
\right ) \eps^2
\right ],
\end{equation}
where we recall that there is no degeneracy splitting in this instance.

\subsubsection{Friction patterns with $n>1$}
We can perform similar analyses for friction functions $g(x)$ with different wavelengths, using the matrices shown in Section~\ref{sec:app:M}. The inverses of these matrices are of the form $\mmat^{-1} = \mmat{}_\text{inv}^{(0)} + M_\text{inv}^{(\eps)}$, with $\mmat{}_\text{inv}^{(0)}$ diagonal and independent of \eps{}, and with $M_\text{inv}^{(\eps)}$ to lowest order of the form:

\begin{align}
M_\text{inv}^{(\eps),n=1}&=\begin{pmatrix}
        \grey{\bigo(\eps^2)} & \bigo(\eps) & \grey{\bigo(\eps^2)} & \grey{\bigo(\eps^3)} & \grey{\bigo(\eps^4)} \\
        \bigo(\eps) & \grey{\bigo(\eps^2)} & \bigo(\eps) & \grey{\bigo(\eps^2)} & \grey{\bigo(\eps^3)} \\
        \grey{\bigo(\eps^2)} & \bigo(\eps) & \grey{\bigo(\eps^2)} & \bigo(\eps) & \grey{\bigo(\eps^2)} \\
        \grey{\bigo(\eps^3)} & \grey{\bigo(\eps^2)} & \bigo(\eps)  & \grey{\bigo(\eps^2)} & \bigo(\eps) \\
        \grey{\bigo(\eps^4)} & \grey{\bigo(\eps^3)} & \grey{\bigo(\eps^2)} & \bigo(\eps) & \grey{\bigo(\eps^2)}
        \end{pmatrix}\\[20pt]
M_\text{inv}^{(\eps),n=2}&=
\begin{pmatrix}
        \grey{\bigo(\eps^2)} & \grey{0} & \bigo(\eps) & \grey{0} & \grey{\bigo(\eps^2)} \\
        \grey{0} & \grey{\bigo(\eps^2)} & \grey{0} & {\color{blue}\bigo(\eps)} & \grey{0} \\
        \bigo(\eps) & \grey{0} & \grey{\bigo(\eps^2)} & \grey{0} & \bigo(\eps) \\
        \grey{0} & {\color{blue}\bigo(\eps)} & \grey{0}  & \grey{\bigo(\eps^2)} & \grey{0} \\
        \grey{\bigo(\eps^2)} & \grey{0} & \bigo(\eps) & \grey{0} & \grey{\bigo(\eps^2)}
\end{pmatrix}\\[20pt]
M_\text{inv}^{(\eps),n=3}&=\begin{pmatrix}
        \grey{\bigo(\eps^2)} & \grey{0} & \grey{0} & \bigo(\eps) & \grey{0} \\
        \grey{0} & \grey{\bigo(\eps^2)} & \grey{0} & \grey{0} & \bigo(\eps) \\
        \grey{0} & \grey{0} & \grey{\bigo(\eps^2)} & \grey{0} & \grey{0} \\
        \bigo(\eps) & \grey{0} & \grey{0}  & \grey{\bigo(\eps^2)} & \grey{0} \\
        \grey{0} & \bigo(\eps) & \grey{0} & \grey{0} & \grey{\bigo(\eps^2)}
        \end{pmatrix}\\[20pt]
M_\text{inv}^{(\eps),n=4}&=
\begin{pmatrix}
        \grey{\bigo(\eps^2)} & \grey{0} & \grey{0} & \grey{0} & {\color{blue}\bigo(\eps)} \\
        \grey{0} & \grey{\bigo(\eps^2)} & \grey{0} & \grey{0} & \grey{0} \\
        \grey{0} & \grey{0} & \grey{\bigo(\eps^2)} & \grey{0} & \grey{0} \\
        \grey{0} & \grey{0} & \grey{0} & \grey{\bigo(\eps^2)} & \grey{0} \\
        {\color{blue}\bigo(\eps)} & \grey{0} & \grey{0} & \grey{0} & \grey{\bigo(\eps^2)}
\end{pmatrix}
\end{align}
etc. The inverse matrix entries with $\bigo(\eps)$ (shown in black/blue, highlighted against all other entries which are in gray) indicate the dominant couplings in the expansion. For $n=2$ and $n=4$, these entries line up with corners of a truncated $3\times 3$ or $5\times 5$ matrix. This is the mathematical explanation of the optimal coupling between unstable modes with wavenumber~$m$ and friction patterning with wavenumber $n=2m$, as illustrated in the main text in Fig~\ref{fig:v_affected_by_fric_wavelength}. When this is the case, the $\mathcal{O}(\eps)$ corrections act directly on the degenerate subspace of the mode $m$, and no higher-order expansion is required to obtain the first correction to the Jacobian. Following essentially identical working to the previous section, we can find the effective Jacobian for the $|m|=1$ degenerate subspace for a friction pattern with $n=2$. This is of the form, to first order in \eps{}:
\begin{align}\label{eq:jachatn2}
\hat{\mathcal{J}}_1^{n=2} &= 
-\left[r+k_1^2(1-\alpha\beta\pe\,f'(1))\right]\mathbb{I}
+\eps k_1^2\frac{\alpha\beta^2}{2}\pe\,f'(1) 
\begin{pmatrix}
0 & -1\\
-1 & 0
\end{pmatrix}\\
&=:\hat{\mathcal{J}}^{n=2,(0)}_{1}+\eps \hat{\mathcal{J}}^{n=2,(1)}_{1},
\end{align}
which leads to eigenvalue corrections:
\begin{equation}
\hat{\lambda}_1^{n=2} = \lambda_1^{n=2,(0)} + \eps \lambda^{n=2,(1)}_1 = - r - k_1^2 + \pe f'(1) k_1^2 \alpha 
\left [ 
\beta_1 \pm \beta_1^2 \left ( \frac{\eps}{2}\right )
\right ].
\end{equation}

\begin{figure}
    \centering
    \includegraphics{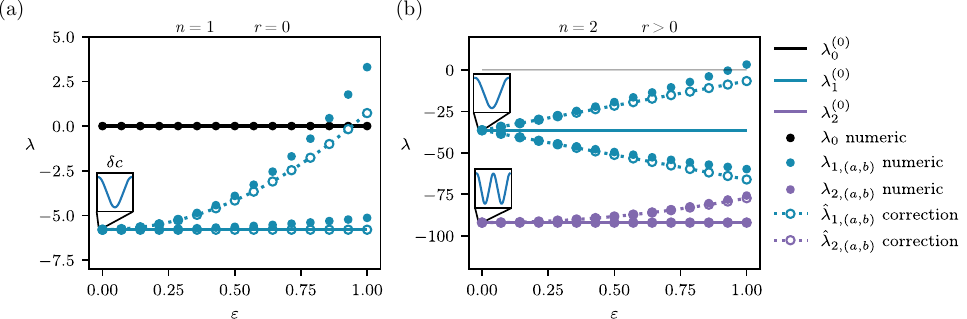}
    \caption{Spectrum of \amat{} for two different parameter sets. (a) $n=1$, $\pe{}=220$, $\alpha = 0.2$, $r=0$. (b) $n=2, \pe{}=1000, \alpha = 0.2, r=150$. Points show numerically calculated eigenvalues. Full lines show the eigenvalues for $\eps=0$. Dashed lines show lowest-order approximations to the eigenvalue corrections using the perturbation method discussed above. Insets show concentration eigenmodes at the indicated locations. }
    \label{fig:app:spectrum}
\end{figure}

This first-order correction is shown in Fig.~\ref{fig:app:spectrum}b. The $|m|=1$ eigenvalues for homogeneous friction are shown in solid blue lines, with blue points showing numerical calculation of the eigenvalues of the full, non-truncated Jacobian matrix for $\eps>0$. Dashed lines show the first-order correction described above. Purple lines and points show similar corrections for the $|m|=2$ eigenvalues for this $n=2$ friction pattern. Not only does inhomogeneous friction promote the growth of unstable modes, but the different plots in Fig~\ref{fig:app:spectrum} illustrate the delicate interplay between the intrinsic length scale of instabilities and the length scale of friction patterning. 

\end{document}